\documentclass[12pt]{article}
\usepackage[top=1in, bottom=1in, left=1in, right=1in]{geometry}
\usepackage{amsmath}
\usepackage{amsfonts}
\usepackage{hyperref}
\usepackage{xcolor}
\usepackage{booktabs}
\usepackage{bm}
\usepackage{graphicx} 
\usepackage{subcaption}
\usepackage{authblk} 
\usepackage{parskip}
\usepackage{natbib}

\newenvironment{widetabular}[1]{%
  \begin{tabular*}{\textwidth}{@{\extracolsep{\fill}}#1}%
}{%
  \end{tabular*}%
}

\definecolor{linkcolor}{RGB}{0, 90, 160}
\hypersetup{
    colorlinks=true,
    linkcolor=linkcolor,
    urlcolor=linkcolor,
    citecolor=linkcolor
}

\title{A Unified Server Quality Metric for Tennis}
\author[1]{Aiwen Li}
\author[1]{Amrita Balajee}
\author[2]{Harry Wieand}
\author[1]{Jonathan Pipping-Gam\'on}
\affil[1]{University of Pennsylvania}
\affil[2]{Boston University Academy}
\date{\today}

\begin{document}

\maketitle

\begin{abstract}
\textcolor{black}{Traditional tennis rating systems (e.g., Elo) summarize overall player strength but do not isolate the independent value of serving. Using point-by-point data from Wimbledon and the U.S.\ Open, we develop serve-specific player metrics that separate serving quality from return ability and other latent factors. For each tournament and gender, we fit logistic mixed-effects models of point outcomes using serve speed, speed variability, and placement features, with crossed server and returner random intercepts to capture unobserved player strengths. From these models we derive Server Quality Scores (SQS): partially pooled, opponent-adjusted estimates of players' serving impact. In out-of-sample evaluation, SQS aligns more strongly with serve efficiency---the probability of winning points within three shots---than weighted Elo. We further benchmark SQS against task-aligned serve-stat baselines and model ablations, quantifying the incremental value of serve features and partial pooling. Associations with overall serve win percentage are smaller and dataset-dependent, and neither SQS nor weighted Elo consistently dominates that outcome. Overall, SQS is best interpreted as a measure of serve-induced short-point advantage (serve quality plus early-point conversion), complementing holistic ratings with actionable insight for coaching, forecasting, and player evaluation.}
\end{abstract}

\section{Introduction}\label{sec:intro}

Current tennis rating systems, such as Elo, estimate player strength from match-level outcomes.
While effective for forecasting, these models compress the complexity of play into a single number and do not isolate the value of specific skills.
In particular, the serve---the only shot completely under a player's control---is treated implicitly rather than modeled directly.
In practice, practitioners often summarize serving performance with simple statistics such as ace rate, first-serve-in percentage, and first-serve-win percentage.
These summaries are easy to compute, but each reflects only a narrow slice of serve quality and can mask the mechanisms that make a serve effective.
Moreover, they conflate distinct dimensions of serving (pace, placement, and variability) and are sensitive to opponent quality and point context.
Taken together, these limitations motivate a holistic, point-level measure of serving performance independent of other skills.

\paragraph{Our Contribution:} This paper establishes a framework to estimate a serve-specific player metric using generalized linear mixed models (GLMMs).
We call this metric a player's \textit{Server Quality Score} (SQS).
SQS captures both measured skill derived from serve features (average speed, speed variability, and location characteristics) and unmeasured server effects modeled through player-specific random intercepts.
To distinguish between aggressive and defensive serving contexts, we fit separate models for first and second serves.
We treat SQS as a two-dimensional metric, reporting separate first-serve and second-serve scores.

Using point-by-point data from Wimbledon and the U.S. Open (2018--2019 and 2021--2024), we benchmark these server metrics against weighted Elo (wElo).
\textcolor{black}{We also compare against task-aligned baselines (standard serve statistics, random-effects-only GLMM scores, and fixed-effects-only scores) to test whether SQS adds information beyond simpler serve summaries and model ablations.}

\paragraph{Organization:} Section~\ref{sec:lit} reviews related work.
Section~\ref{sec:methods} describes our methodology with data preparation, mixed effects models, and construction of SQS.
Section~\ref{sec:results} discusses out-of-sample testing results and benchmarks SQS against wElo.
\textcolor{black}{Additional task-aligned baseline analyses, temporal validation, and point-level robustness checks are reported in the appendix.}
We conclude with a discussion in Section~\ref{sec:discussion}.

\section{Related Work}\label{sec:lit}
\subsection{Match-Level Ratings and Forecasting}\label{sec:lit:elo}
Elo-style ratings, adapted from chess and widely used in tennis, estimate a player's strength from match results.
These ratings have consistently been effective for forecasting match winners using large datasets (\cite{klaassen_forecasting_2003}; \cite{kovalchik_searching_2016}).
A recent extension, weighted Elo (wElo), incorporates margins of victory and has been shown to outperform standard Elo and other common baselines, including in value-betting applications (\cite{angelini_weighted_2022}).

Beyond pre-match ratings, other studies combine in-play information to improve point-by-point forecasting.
For example, \cite{kovalchik_calibration_2019} show that starting from a rating-based prior and updating dynamically during a match improves win probability estimates.

Alternative approaches model tennis as a hierarchical Markov process, where point-level win probabilities map recursively to game, set, and match win probabilities.
Early work studied whether points are independent and identically distributed (IID), finding that this assumption is not always true due to changes in momentum and pressure (\cite{klaassen_are_2001}).

Despite these limitations, models under the IID assumption still provide a useful baseline.
\cite{omalley_probability_2008} derives exact game, set, and match win probabilities under an IID assumption and studies how these probabilities change with serve and return strength.
Later studies relax the IID assumption by using state-dependent dynamics that better predict live win probabilities (\cite{klaassen_forecasting_2003}; \cite{newton_monte_2009}).

While these ratings and match models are valuable for forecasting, they intentionally aggregate performance across all phases of play.
This motivates serve-specific modeling that can complement Elo-style summaries when the goal is interpretation and skill isolation rather than match prediction.

\subsection{Point-Level Models and Serve Win Probabilities}\label{sec:lit:serveprob}
A central part of point-based tennis models is a player's ``serve win probability'', defined as the probability of winning a point as the server.
These probabilities are often split between first and second serves, and they can be adjusted based on surface or match context.

Early work by \cite{barnett_combining_2005} computes match-specific serve probabilities by combining each player's historical serve and return performance with opponent strength.
Subsequent work refines this approach by adding surface adjustments, shrinkage for small samples, and common-opponent comparisons.
More recently, \cite{gollub_forecasting_2021} improves estimates of serve win probabilities by combining broader Elo-type player ratings.

However, serve win probabilities in point-based models are typically treated as context-dependent inputs for predicting match outcomes rather than as standalone measures of serving quality. 
Our goal is instead to estimate an interpretable server metric derived from serve characteristics while accounting for unobserved, player-specific effects via partial pooling.

\subsection{Gap: Isolating Serve Quality}\label{sec:lit:open}
Despite frequent use of serving variables in both match- and point-level models, there is still no widely adopted metric that isolates the serve as an independent component of player quality.
In practice, separating the serve's contribution from other aspects of play improves interpretability and supports decision-making.
It clarifies how much advantage comes from the serve itself and informs coaching strategies that are specific to serving and returning.

Motivated by this gap, we introduce a framework for serve-specific player metrics focused solely on serving performance.
Section~\ref{sec:methods} describes the methodology used to construct these metrics.

\section{Methodology}\label{sec:methods}

\subsection{Data and Feature Construction}\label{sec:methods:data}

We use publicly available point-by-point data for singles matches at the Wimbledon and U.S. Open tournaments from \href{https://github.com/JeffSackmann/tennis_slam_pointbypoint}{Jeff Sackmann's public tennis database}.
We pool six seasons of data (2018--2019 and 2021--2024) into four groups: Wimbledon men's singles, Wimbledon women's singles, U.S. Open men's singles, and U.S. Open women's singles (excluding 2020 due to COVID-19--related disruptions).
All analyses are performed separately for each tournament and gender.

Within each dataset, matches are randomly split into training (80\%) and testing (20\%) sets \emph{within each year}.
This ensures that all points from a given match are assigned to the same split and that each season contributes to both training and testing, avoiding leakage from within-match dependence while balancing year-to-year variation.
Server Quality Scores (SQS) are constructed using the training set, and the testing set is reserved for out-of-sample evaluation.
{\color{black}As a complementary check, we also evaluate SQS under a temporal split (train on 2018--2022, test on 2023--2024); these results appear in Appendix~\ref{app:temporal}.\normalcolor}

We begin by cleaning the raw point-level data, applying the following steps:
\begin{itemize}
  \item Removing serves with missing location information.
  \item Restricting the data to valid first and second serves.
  \item Excluding serves recorded with zero speed (due to faults).
  \item Defining serve location using two categorical variables, \texttt{ServeWidth} and \texttt{ServeDepth}, and assigning each serve to a discrete width--depth location bin.
\end{itemize}

\[
\texttt{location\_bin} = (\texttt{ServeWidth},\ \texttt{ServeDepth}).
\]
For each server $j$ and serve type $s \in \{1,2\}$ (first or second), we summarize serve behavior using a set of interpretable features:
\begin{itemize}
  \item $n^{(s)}_{j}$: number of serves observed of type $s$.
  \item $\text{avg\_speed}^{(s)}_{j}$: average serve speed (mph),
  \item $\text{sd\_speed}^{(s)}_{j}$: standard deviation of serve speed (a proxy for variability/unpredictability),
  \item $\text{modal\_loc}^{(s)}_{j}$: the modal location bin (one-hot encoded in the model),
  \item $\text{loc\_entropy}^{(s)}_{j}$: location entropy (a proxy for location unpredictability), defined as
\end{itemize}
\[
\text{loc\_entropy}^{(s)}_{j} = -\sum_{b} p^{(s)}_{b,j}\log_2 p^{(s)}_{b,j},
\]
where $p^{(s)}_{b,j}$ is the proportion of serves by player $j$ (of type $s$) that land in location bin $b$.
These features were chosen to balance interpretability and signal.
Average serve speed summarizes pace, $\text{sd\_speed}$ captures pace unpredictability, $\text{modal\_loc}$ captures directional tendencies, and location entropy captures how dispersed a player's placement is across coarse bins.
To ensure that our estimates are reliable, we only keep servers with more than 20 serves of a given type.
Any continuous features are standardized within the dataset (denoted by the superscript $z$), and serve locations are encoded categorically.

{\color{black}Because these predictors summarize serve behavior at the player level, they compress within-player variation across individual points. As a robustness check, we implement alternative models using point-level serve features (serve speed, location bin, and an indicator for whether the serve matches the player's modal location) and report the results in Appendix~\ref{app:point_level}. The alternative results are broadly consistent with the main specification, suggesting that the aggregated server-level model retains most of the predictive signal relevant to serve efficiency while remaining easier to interpret as a stable player-level serve profile.\normalcolor}

Finally, we consider two complementary outcomes: overall point win percentage and \emph{serve efficiency}.
We define an efficient serve as one where the server wins the point within the first three shots (serve, return, and the server's next shot).
\textcolor{black}{This restriction is deliberate: by focusing on the earliest phase of the point---when the server's initial delivery most directly shapes the exchange---serve efficiency acts as a measure of serve-induced short-point advantage rather than overall point wins, which increasingly reflect baseline rally skill and endurance as the point lengthens.
In this sense, serve efficiency captures situations in which the serve generates an immediate advantage (e.g., an ace, forced error, weak return, or short-ball put-away) that the server converts quickly, rather than points in which the serve merely initiates a neutral rally.}

In our analysis, we model the point-level \emph{efficient-serve indicator} (a $0/1$ variable under the three-shot definition above) as the outcome in Section~\ref{sec:methods:glmm}.
These scores are then evaluated using aggregated \textit{server-level} serve efficiency and win percentage from our testing dataset in Section~\ref{sec:methods:oos}.

\subsection{Mixed-Effects Model and Server Quality Score (SQS)}\label{sec:methods:glmm}
Our goal is to quantify server quality in a way that reflects both observable serve characteristics and unobservable player effects.
To do so, we fit separate logistic mixed-effects models for first and second serves to account for different serving contexts.
This separation allows the relationship between serve features and short-point outcomes to differ across serve types.

Each model is fit at the point level, with the outcome being whether the serve was \textit{efficient}, i.e., whether the server won the point within three total shots (as defined in Section~\ref{sec:methods:data}).

For serve type $s\in\{1,2\}$, let $Y^{(s)}_{i,j}$ be the binary indicator that server $j$ wins point $i$ within three shots.
We model this using a logistic mixed-effects framework:
\[
\begin{aligned}
\text{logit}\,\Pr(Y^{(s)}_{i,j}=1) &= \beta^{(s)}_{0}
+ \beta^{(s)}_{1}\,\text{avg\_speed}^{(s,z)}_{j}
+ \beta^{(s)}_{2}\,\text{sd\_speed}^{(s,z)}_{j} \\
&\quad + (\boldsymbol{\beta}^{(s)}_{3})^{\top}\,\mathbb{I}(\text{modal\_loc}^{(s)}_{j})
+ \beta^{(s)}_{4}\,\text{loc\_entropy}^{(s,z)}_{j} \\
&\quad + u^{(s)}_{j} + v^{(s)}_{k}
\end{aligned}
\]
where $u^{(s)}_{j}\sim \mathcal{N}(0,\sigma_s^2)$ is a server-level random intercept and
$v^{(s)}_{k}\sim \mathcal{N}(0,\tau_s^2)$ is a returner-level random intercept (for returner $k$).
The superscript $z$ on speed and entropy features indicates standardization (z-scoring) within the dataset.

The fixed effects quantify the association between observable serve characteristics (pace, pace variability, and placement tendencies) and short-point success.
The returner-specific random intercept $v^{(s)}_{k}$ adjusts for opponent return strength.
Without this term, servers who disproportionately face weaker returners (e.g., due to draw effects) can appear artificially strong.
By including crossed random intercepts for server and returner, the estimated server effect $u^{(s)}_{j}$ is identified net of the average return quality faced, yielding an opponent-adjusted measure of serving impact.

For each serve type, we summarize the fitted model into a single \textit{server quality score} by combining each player's estimated fixed and random effects:
%
\[
\begin{aligned}
\text{SQS}^{(s)}_{j}
&:= \hat{\beta}^{(s)}_{0}
+ \hat{\beta}^{(s)}_{1}\,\text{avg\_speed}^{(s,z)}_{j}
+ \hat{\beta}^{(s)}_{2}\,\text{sd\_speed}^{(s,z)}_{j} \\
&\quad + (\hat{\boldsymbol{\beta}}^{(s)}_{3})^{\top}\,\mathbb{I}(\text{modal\_loc}^{(s)}_{j})
+ \hat{\beta}^{(s)}_{4}\,\text{loc\_entropy}^{(s,z)}_{j}
+ \hat{u}^{(s)}_{j}.
\end{aligned}
\]

We report $(\text{SQS}^{(1)}_{j},\ \text{SQS}^{(2)}_{j})$ as a two-dimensional serving profile.
The first-serve score $\text{SQS}^{(1)}_{j}$ summarizes how much advantage a player's first delivery tends to create in the three-shot sequence, reflecting the payoff of pace, placement, and disguise in more aggressive serving contexts.
The second-serve score $\text{SQS}^{(2)}_{j}$ summarizes effectiveness under the more conservative second-serve regime, where the objective shifts toward limiting opponent aggression and securing a playable first-strike opportunity.

Because the fitted model includes both server and returner effects, we define SQS as the \emph{server-side} linear predictor evaluated at an average returner (i.e., setting $v^{(s)}_{k}=0$).
Equivalently, SQS summarizes the predicted short-point advantage attributable to the server after adjusting for returner strength.
We keep SQS on the log-odds scale of the fitted models, since this is the natural additive scale on which fixed effects and random effects combine. \textcolor{black}{On this scale, differences in SQS correspond to multiplicative changes in the odds of winning the point, so higher values indicate serves that substantially increase the likelihood of winning the rally (e.g., a 0.10 increase in SQS corresponds to an odds ratio of $\exp(0.10)\approx 1.11$).}

\subsection{Out-of-Sample Evaluation and Baselines}\label{sec:methods:oos}

We evaluate Server Quality Scores (SQS) out of sample using the held-out matches in each tournament--gender dataset.
Since SQS is defined at the server-by-serve-type level, we aggregate test-set points by server and compute two server-level outcomes: overall point win percentage and serve efficiency (defined in Section~\ref{sec:methods:data}).
Serve efficiency captures short-point success within three shots, while win percentage summarizes outcomes over all rally lengths:
\[
\widehat{\text{ServeEff}}_j=\frac{\#\{\text{serve points won with }\mathrm{RallyCount}\le 3\}}{\#\{\text{serve points}\}}, \qquad \widehat{\text{WinPct}}_j=\frac{\#\{\text{serve points won}\}}{\#\{\text{serve points}\}}
\]

For each serve type $s\in\{1,2\}$ and each outcome, we model the number of successes for server $j$ as binomial with denominator equal to the number of test-set serves of type $s$.
We then fit binomial GLMs with grouped outcomes on the corresponding serve-type score, using $\text{SQS}^{(1)}_{j}$ for first serves and $\text{SQS}^{(2)}_{j}$ for second serves.
These regressions quantify how SQS relates to short-point success (serve efficiency) versus overall point success (win percentage) in unseen matches.

As a baseline, we repeat the same evaluation using weighted Elo (wElo) in place of SQS.
wElo scores are computed using the R package \texttt{welo} \citep{angelini_weighted_2022}, providing a match-level benchmark against which to assess the incremental information in serve-specific scores.

\textcolor{black}{To address task alignment directly, we also evaluate three additional baseline families: (i) standard serve statistics (ace rate, first-serve points won, unreturned-serve proxy rate, and first-serve-in\%), (ii) a random-effects-only GLMM with server and returner intercepts but no serve covariates, and (iii) a fixed-effects-only score that uses the measured serve-feature component of SQS without the server random effect.}

\textcolor{black}{For comparability, each tournament--gender--serve-type evaluation uses a common complete-case server set across all predictors. Full baseline comparisons are reported in Appendix~\ref{app:task_baselines}.}

\textcolor{black}{In Section~\ref{sec:results}, we report regression coefficients and correlations between each predictor and the observed test-set outcomes.}

\section{Results}\label{sec:results}
We report full out-of-sample results for Wimbledon men's singles in Tables~\ref{tab:wimbledon_males_first} and~\ref{tab:wimbledon_males_second}.
Analogous tables for Wimbledon women's singles and the U.S.\ Open (men and women) appear in Appendix~\ref{app:oos_results}; we reference them here only to summarize cross-dataset patterns. 

\textcolor{black}{Expanded task-aligned baseline comparisons appear in Appendix~\ref{app:task_baselines}. In addition, results from the alternative point-level specification are reported in Appendix~\ref{app:point_level}. These results show similar qualitative patterns, indicating that the server-level specification used in the main analysis captures most of the predictive signal while remaining easier to interpret as a stable player-level serve profile.}

\begin{table}[htbp!]
\centering
\caption{Out-of-sample performance for Wimbledon men's singles (first serves).}
\label{tab:wimbledon_males_first}
\footnotesize
\begin{widetabular}{lccccc}
\toprule
\textbf{Outcome} & \textbf{Predictor} & \textbf{n} & \textbf{Coefficient} & \textbf{$p$-value} & \textbf{Correlation ($r$)} \\
\midrule
Serve efficiency & SQS$_1$ & 62 & 0.240 & $4.6\times10^{-22}$ & 0.667 \\
Serve efficiency & wElo    & 62 & 0.015 & 0.563 & 0.148 \\
Win percentage   & SQS$_1$ & 62 & 0.109 & $1.5\times10^{-5}$ & 0.325 \\
Win percentage   & wElo    & 62 & 0.030 & 0.256 & 0.136 \\
\bottomrule
\end{widetabular}
\end{table}

\begin{table}[htbp!]
\centering
\caption{Out-of-sample performance for Wimbledon men's singles (second serves).}
\label{tab:wimbledon_males_second}
\footnotesize
\begin{widetabular}{lccccc}
\toprule
\textbf{Outcome} & \textbf{Predictor} & \textbf{n} & \textbf{Coefficient} & \textbf{$p$-value} & \textbf{Correlation ($r$)} \\
\midrule
Serve efficiency & SQS$_2$ & 60 & 0.127 & 0.00072 & 0.232 \\
Serve efficiency & wElo    & 60 & 0.017 & 0.663 & 0.090 \\
Win percentage   & SQS$_2$ & 60 & -0.064 & 0.064 & -0.251 \\
Win percentage   & wElo    & 60 & 0.014 & 0.692 & 0.056 \\
\bottomrule
\end{widetabular}
\end{table}

\subsection{Serve Efficiency vs. Point Win Percentage}\label{sec:results:effwin}
Across tournaments and genders, SQS aligns most closely with serve efficiency, our serve-proximal target.
On first serves, SQS is positively associated with serve efficiency in all four datasets and is especially strong at Wimbledon (e.g., $r=0.667$ for Wimbledon men and $r=0.564$ for Wimbledon women), with more modest associations at the U.S.\ Open ($r\approx 0.28$ for men and $r\approx 0.24$ for women).
The larger Wimbledon associations are consistent with surface effects: grass-court match-play features shorter rallies and a more pronounced serve/return advantage, so a three-shot outcome concentrates more of the point's signal in the opening exchange \citep{fitzpatrick2019}.
On second serves, SQS--serve efficiency associations are smaller and less stable, but remain positive in three of four datasets (including Wimbledon men: $r=0.232$), with U.S.\ Open men as an exception ($r\approx -0.08$).
In contrast, wElo exhibits weak or negative correlations with serve efficiency across datasets, consistent with a match-level rating that is not designed to isolate serve-driven short-point advantage.

\textcolor{black}{Against the additional task-aligned baselines (Appendix~\ref{app:task_baselines}), SQS remains strongest on average for first-serve efficiency ($\bar r=0.439$), ahead of fixed-effects-only ($0.423$), ace rate ($0.406$), and random-effects-only ($0.388$), while wElo is weak on this serve-proximal target ($\bar r=-0.067$). At the dataset level, SQS is the top first-serve efficiency predictor in three of four splits.}

\textcolor{black}{The ablation comparisons support incremental value from both model components. Relative to the random-effects-only model, SQS improves first-serve efficiency correlations in all four datasets, indicating added signal from serve-feature covariates. Relative to fixed-effects-only, SQS is higher in three of four datasets, suggesting that partial pooling via server random effects contributes in most settings.}

Associations with overall point win percentage are generally weaker and more variable for both ratings.
On first serves, SQS correlations remain positive across datasets, while wElo is small and sometimes negative.
On second serves, both predictors show mixed performance, suggesting that overall point outcomes under second-serve conditions depend strongly on broader skills and contextual factors beyond the serve alone.
\textcolor{black}{Task-aligned baseline comparisons for win percentage (Appendix~\ref{app:task_baselines}) reinforce this pattern. On first serves, mean correlations are similar across several predictors (SQS: $\bar r=0.259$, fixed-effects-only: $0.261$, ace rate: $0.240$), so no single metric clearly dominates. On second serves, SQS is weaker on average ($\bar r=-0.068$), while wElo is relatively stronger ($\bar r=0.111$), consistent with win percentage reflecting broader point-winning skill beyond serve-proximal effects.}
This outcome-specific separation is consistent with SQS isolating serve-proximal impact: it aligns most closely with a target designed to concentrate signal in the earliest shots, and less closely with outcomes increasingly driven by longer-rally dynamics.
Player ranking tables for first and second serves are provided in Appendix~\ref{app:rankings}.

\subsection{First vs. Second Serves}\label{sec:results:firstsecond}
Stratifying by serve type clarifies how serving context mediates the relationship between ratings and outcomes.
On first serves, servers have the greatest opportunity to create immediate leverage; correspondingly, SQS shows its strongest and most reliable associations with the serve-efficiency target.
On second serves, where pace is reduced and the returner typically sees more playable deliveries, efficiency and win-percentage outcomes depend more heavily on the ensuing exchange, making relationships with any serve-only summary less stable across datasets.

\subsection{Implications}\label{sec:results:implications}
Overall, the out-of-sample results support interpreting SQS as a serve-focused metric.
Across datasets, SQS is most informative for explaining variation in serve efficiency---an outcome designed to concentrate signal in the serve--return--first-strike phase---whereas associations with overall point win percentage are weaker and more heterogeneous, especially on second serves where immediate serve leverage is reduced.
Although serve efficiency still reflects the return and the server's next shot, restricting attention to a three-shot window reduces the influence of longer-rally dynamics. This provides empirical evidence that SQS captures a distinct, serve-proximal component of performance.

\section{Discussion}\label{sec:discussion}

{\color{black}\subsection{Comparison with Prior Work}\label{sec:discussion:comparison}\normalcolor}
\textcolor{black}{These results are consistent with prior tennis-rating research showing that match-level metrics such as Elo-type ratings are effective summaries of overall performance, while extending that literature by isolating a serve-proximal component. Relative to prior work on serve win probabilities as inputs to match forecasting, SQS is designed as an interpretable player metric rather than a match predictor. The direct comparison with weighted Elo and the additional win-percentage analyses show this distinction empirically: SQS is strongest for short-point serve efficiency, whereas weighted Elo is often more aligned with broader point outcomes, especially on second serves.}

{\color{black}\subsection{Conclusions and Practical Applications}\label{sec:conclusions}\normalcolor}
This paper proposes a serve-specific framework for evaluating tennis performance using point-level data.
We introduce a Server Quality Score (SQS), estimated via logistic mixed-effects models, that summarizes how effectively a player's serve converts the opening exchange into an early-point advantage.

Across tournaments and genders, out-of-sample evaluation supports the intended interpretation of SQS as a serve-proximal metric.
SQS aligns most strongly with serve efficiency, while relationships with overall point win percentage are smaller and more heterogeneous: especially on second serves where immediate serve leverage is reduced.
Weighted Elo (wElo), by contrast, reflects broader point-winning ability and in some settings aligns more closely with win percentage; however, neither rating uniformly dominates for win percentage across all datasets.
Taken together, these results provide empirical evidence that SQS captures a distinct, serve-driven component of performance that complements holistic match-level ratings.

\subsection{Limitations and Future Work}\label{sec:limitations}
Several limitations suggest directions for extending the SQS framework.
We adjust for average return strength via a returner random intercept, but we do not model server--returner interaction effects (matchup-specific styles) or contextual shifts in return positioning.
First, the model omits contextual factors such as score state, point importance, and fatigue.
Because serve selection and execution change under pressure, incorporating context could help distinguish underlying serve quality from strategic adaptation within matches.

Second, our placement features (modal location and location entropy) provide interpretable summaries but compress richer spatial structure.
Future work could incorporate continuous serve coordinates, spatial smoothing, or probabilistic location models to better represent placement strategies and their interaction with pace.

Third, data quality limits the characterization of risk.
In particular, serves with recorded speed of zero (typically faults or missing tracking) were excluded, potentially omitting information about second-serve safety and risk--reward tradeoffs.
More complete tracking of serve attempts and faults would enable models that jointly capture placement, speed, and error propensity.

{\color{black}Finally, our analysis is conducted separately by tournament and gender, which means the resulting SQS values are not directly comparable across tournaments or surfaces. For example, faster courts such as grass typically amplify serve advantage relative to slower hard or clay courts. In this study, we partially address this concern by estimating separate models for each tournament. The top-ranked servers in Appendix~\ref{app:rankings} are broadly consistent across the Wimbledon and U.S. Open datasets, suggesting that the metric captures stable aspects of serving ability despite surface differences. An extension to this would be to estimate a hierarchical model that pools information across tournaments and seasons while allowing surface-specific effects. This would enable direct comparisons of serving quality across surfaces and competitions.\normalcolor}

Together, these extensions provide a path toward richer serve-specific metrics through contextual modeling and improved spatial representation, while preserving the interpretability and portability of SQS.

\section{Reproducibility}\label{sec:reproducibility}
All code is available on the project's \href{https://github.com/WhartonSABI/server-quality}{GitHub repository}.

\section{Acknowledgments}\label{sec:acknowledgments}
The authors thank Professor Abraham J. Wyner and Tianshu Feng for helpful feedback, Audrey Kuan for her contributions during the summer, and the Wharton Sports Analytics and Business Initiative (WSABI) for supporting this research.

\bibliographystyle{apalike}
\bibliography{references}

\clearpage

\appendix
\begin{center}
\Large\bfseries Appendix
\end{center}

\section{Additional Out-of-Sample Results}\label{app:oos_results}

\begin{table}[htbp!]
\centering
\caption{Out-of-sample performance across datasets (first serves).}
\label{tab:oos_first}
\footnotesize
\begin{widetabular}{llccccc}
\toprule
\textbf{Dataset} & \textbf{Outcome} & \textbf{Predictor} & \textbf{n} & \textbf{Coefficient} & \textbf{$p$-value} & \textbf{Correlation ($r$)} \\
\midrule
Wimbledon women & Serve efficiency & SQS$_1$ & 65 & 0.301 & $1.1\times10^{-22}$ & 0.564 \\
Wimbledon women & Serve efficiency & wElo    & 65 & $-0.020$ & 0.511 & $-0.102$ \\
Wimbledon women & Win percentage   & SQS$_1$ & 65 & 0.134 & $8.1\times10^{-6}$ & 0.447 \\
Wimbledon women & Win percentage   & wElo    & 65 & $-0.015$ & 0.625 & $-0.123$ \\
\midrule
U.S.\ Open men    & Serve efficiency & SQS$_1$ & 86 & 0.098 & $8.9\times10^{-6}$ & 0.283 \\
U.S.\ Open men    & Serve efficiency & wElo    & 86 & $-0.074$ & $3.5\times10^{-4}$ & $-0.245$ \\
U.S.\ Open men    & Win percentage   & SQS$_1$ & 86 & 0.061 & 0.0064 & 0.164 \\
U.S.\ Open men    & Win percentage   & wElo    & 86 & 0.044 & 0.038 & 0.011 \\
\midrule
U.S.\ Open women  & Serve efficiency & SQS$_1$ & 92 & 0.123 & $2.4\times10^{-6}$ & 0.243 \\
U.S.\ Open women  & Serve efficiency & wElo    & 92 & 0.004 & 0.866 & $-0.070$ \\
U.S.\ Open women  & Win percentage   & SQS$_1$ & 92 & 0.046 & 0.071 & 0.099 \\
U.S.\ Open women  & Win percentage   & wElo    & 92 & 0.036 & 0.154 & 0.032 \\
\bottomrule
\end{widetabular}
\end{table}

\begin{table}[htbp!]
\centering
\caption{Out-of-sample performance across datasets (second serves).}
\label{tab:oos_second}
\footnotesize
\begin{widetabular}{llccccc}
\toprule
\textbf{Dataset} & \textbf{Outcome} & \textbf{Predictor} & \textbf{n} & \textbf{Coefficient} & \textbf{$p$-value} & \textbf{Correlation ($r$)} \\
\midrule
Wimbledon women & Serve efficiency & SQS$_2$ & 45 & 0.129 & 0.0061 & 0.331 \\
Wimbledon women & Serve efficiency & wElo    & 45 & 0.033 & 0.531 & 0.121 \\
Wimbledon women & Win percentage   & SQS$_2$ & 45 & 0.048 & 0.254 & 0.215 \\
Wimbledon women & Win percentage   & wElo    & 45 & 0.100 & 0.034 & 0.287 \\
\midrule
U.S.\ Open men    & Serve efficiency & SQS$_2$ & 82 & 0.009 & 0.782 & $-0.081$ \\
U.S.\ Open men    & Serve efficiency & wElo    & 82 & $-0.154$ & $3.9\times10^{-7}$ & $-0.321$ \\
U.S.\ Open men    & Win percentage   & SQS$_2$ & 82 & $-0.036$ & 0.196 & $-0.139$ \\
U.S.\ Open men    & Win percentage   & wElo    & 82 & 0.070 & 0.0091 & 0.168 \\
\midrule
U.S.\ Open women  & Serve efficiency & SQS$_2$ & 63 & 0.092 & 0.021 & 0.181 \\
U.S.\ Open women  & Serve efficiency & wElo    & 63 & $-0.121$ & 0.0050 & $-0.279$ \\
U.S.\ Open women  & Win percentage   & SQS$_2$ & 63 & $-0.004$ & 0.921 & $-0.096$ \\
U.S.\ Open women  & Win percentage   & wElo    & 63 & $-0.022$ & 0.574 & $-0.067$ \\
\bottomrule
\end{widetabular}
\end{table}

\clearpage

{\color{black}%
\section{Task-Aligned Baseline Comparisons}\label{app:task_baselines}

To supplement the main SQS--wElo comparison, this appendix reports task-aligned baselines requested in review: standard serve statistics, a random-effects-only GLMM score (server/returner intercepts only), and a fixed-effects-only score (measured serve-feature component without server random effects). All values below use the within-year random split and the common complete-case server set per dataset and serve type.

\begin{table}[htbp!]
\centering\color{black}
\caption{Serve-efficiency correlations by predictor, averaged across datasets (random split).}
\label{tab:task_baselines_mean}
\footnotesize
\begin{widetabular}{lcc}
\toprule
\textbf{Predictor} & \textbf{First-serve mean $r$} & \textbf{Second-serve mean $r$} \\
\midrule
SQS & 0.439 & 0.165 \\
FE-only score & 0.423 & 0.129 \\
Ace rate & 0.406 & 0.107 \\
RE-only GLMM & 0.388 & 0.143 \\
Unreturned rate & 0.318 & 0.047 \\
First-serve points won & 0.245 & 0.066 \\
First-serve-in \% & $-0.030$ & $-0.012$ \\
wElo & $-0.067$ & $-0.097$ \\
\bottomrule
\end{widetabular}
\end{table}

\begin{table}[htbp!]
\centering\color{black}
\caption{First-serve efficiency correlations by dataset for key baseline families (random split).}
\label{tab:task_baselines_first_dataset}
\footnotesize
\begin{widetabular}{lccccc}
\toprule
\textbf{Dataset} & \textbf{SQS} & \textbf{Best serve stat} & \textbf{RE-only GLMM} & \textbf{FE-only score} & \textbf{wElo} \\
\midrule
Wimbledon men   & 0.667 & 0.662 (ace rate) & 0.639 & 0.635 & 0.148 \\
Wimbledon women & 0.564 & 0.463 (ace rate) & 0.549 & 0.502 & $-0.102$ \\
U.S.\ Open men    & 0.283 & 0.262 (ace rate) & 0.261 & 0.263 & $-0.245$ \\
U.S.\ Open women  & 0.243 & 0.237 (ace rate) & 0.102 & 0.292 & $-0.070$ \\
\bottomrule
\end{widetabular}
\end{table}
\begin{table}[htbp!]
\centering\color{black}
\caption{Win-percentage correlations by predictor, averaged across datasets (random split).}
\label{tab:task_baselines_win_mean}
\footnotesize
\begin{widetabular}{lcc}
\toprule
\textbf{Predictor} & \textbf{First-serve mean $r$} & \textbf{Second-serve mean $r$} \\
\midrule
FE-only score & 0.261 & $-0.059$ \\
SQS & 0.259 & $-0.068$ \\
Ace rate & 0.240 & $-0.015$ \\
First-serve points won & 0.231 & $-0.020$ \\
RE-only GLMM & 0.223 & $-0.088$ \\
Unreturned rate & 0.103 & $-0.152$ \\
First-serve-in \% & 0.050 & 0.061 \\
wElo & 0.014 & 0.111 \\
\bottomrule
\end{widetabular}
\end{table}

\begin{table}[htbp!]
\centering\color{black}
\caption{Second-serve win-percentage correlations by dataset for key baseline families (random split).}
\label{tab:task_baselines_win_second_dataset}
\footnotesize
\setlength{\tabcolsep}{2.5pt}
\begin{widetabular}{lccccc}
\toprule
\textbf{Dataset} & \textbf{SQS} & \textbf{Best serve stat} & \textbf{RE-only GLMM} & \textbf{FE-only score} & \textbf{wElo} \\
\midrule
Wimbledon men   & $-0.251$ & 0.050 (1st-srv pts won) & $-0.174$ & $-0.262$ & 0.056 \\
Wimbledon women & 0.215 & 0.158 (1st-srv pts won) & 0.000 & 0.215 & 0.287 \\
U.S.\ Open men    & $-0.139$ & $-0.011$ (ace rate) & $-0.052$ & $-0.203$ & 0.168 \\
U.S.\ Open women  & $-0.096$ & 0.254 (first-serve-in \%) & $-0.127$ & 0.015 & $-0.067$ \\
\bottomrule
\end{widetabular}
\end{table}

The task-aligned baseline results are broadly consistent with the main analysis. For the serve-proximal target (serve efficiency), SQS is strongest on average for first serves and is top in three of four dataset splits, indicating that it captures short-point serve impact better than match-level wElo and most serve-stat baselines. The ablation rows also support incremental value from both components of SQS: compared with random-effects-only, adding serve-feature covariates improves alignment in all first-serve datasets; compared with fixed-effects-only, adding partial pooling improves alignment in most settings. For overall win percentage, predictor performance is flatter and more heterogeneous---especially on second serves, where wElo is often relatively stronger---consistent with this outcome reflecting broader point-construction skill beyond serve-proximal effects.
}\normalcolor

\clearpage

{\color{black}%
\section{Temporal Validation}\label{app:temporal}
The results in Section~\ref{sec:results} and Appendix~\ref{app:oos_results} use an 80/20 random split within each year.
To assess whether SQS generalizes across time, we conduct an additional out-of-time validation:
models are trained on 2018--2019 and 2021--2022 data and evaluated on held-out 2023--2024 data, with no temporal overlap between training and testing.
This design tests whether serve profiles estimated from earlier seasons remain predictive of future performance.

\begin{table}[htbp!]
\centering\color{black}
\caption{Temporal validation: serve efficiency (train 2018--2022, test 2023--2024).}
\label{tab:temporal_eff}
\footnotesize
\begin{widetabular}{llcccc}
\toprule
\textbf{Dataset} & \textbf{Predictor} & \textbf{n} & \textbf{Coefficient} & \textbf{$p$-value} & \textbf{Correlation ($r$)} \\
\midrule
\multicolumn{6}{l}{\textit{First serve}} \\
\midrule
Wimbledon men   & SQS$_1$ & 49 & 0.184 & $5.3\times10^{-15}$ & 0.649 \\
Wimbledon men   & wElo    & 49 & 0.015 & 0.512 & 0.010 \\
Wimbledon women & SQS$_1$ & 63 & 0.151 & $7.6\times10^{-9}$ & 0.365 \\
Wimbledon women & wElo    & 63 & 0.063 & 0.009 & 0.087 \\
U.S.\ Open men    & SQS$_1$ & 80 & 0.088 & $1.1\times10^{-5}$ & 0.297 \\
U.S.\ Open men    & wElo    & 80 & 0.052 & $2.6\times10^{-4}$ & 0.247 \\
U.S.\ Open women  & SQS$_1$ & 85 & 0.224 & $6.3\times10^{-22}$ & 0.578 \\
U.S.\ Open women  & wElo    & 85 & 0.150 & $1.9\times10^{-19}$ & 0.459 \\
\midrule
\multicolumn{6}{l}{\textit{Second serve}} \\
\midrule
Wimbledon men   & SQS$_2$ & 47 & 0.112 & 0.005 & 0.535 \\
Wimbledon men   & wElo    & 47 & 0.023 & 0.485 & 0.024 \\
Wimbledon women & SQS$_2$ & 45 & 0.087 & 0.027 & 0.347 \\
Wimbledon women & wElo    & 45 & 0.058 & 0.160 & 0.312 \\
U.S.\ Open men    & SQS$_2$ & 78 & $-0.007$ & 0.783 & $-0.158$ \\
U.S.\ Open men    & wElo    & 78 & $-0.016$ & 0.442 & $-0.073$ \\
U.S.\ Open women  & SQS$_2$ & 66 & 0.055 & 0.076 & 0.114 \\
U.S.\ Open women  & wElo    & 66 & 0.042 & 0.115 & 0.145 \\
\bottomrule
\end{widetabular}
\end{table}

\begin{table}[htbp!]
\centering\color{black}
\caption{Temporal validation: win percentage (train 2018--2022, test 2023--2024).}
\label{tab:temporal_win}
\footnotesize
\begin{widetabular}{llcccc}
\toprule
\textbf{Dataset} & \textbf{Predictor} & \textbf{n} & \textbf{Coefficient} & \textbf{$p$-value} & \textbf{Correlation ($r$)} \\
\midrule
\multicolumn{6}{l}{\textit{First serve}} \\
\midrule
Wimbledon men   & SQS$_1$ & 49 & 0.120 & $5.6\times10^{-7}$ & 0.548 \\
Wimbledon men   & wElo    & 49 & 0.063 & 0.006 & 0.229 \\
Wimbledon women & SQS$_1$ & 63 & 0.085 & 0.001 & 0.218 \\
Wimbledon women & wElo    & 63 & 0.075 & 0.002 & 0.194 \\
U.S.\ Open men    & SQS$_1$ & 80 & 0.052 & 0.011 & 0.263 \\
U.S.\ Open men    & wElo    & 80 & 0.102 & $5.1\times10^{-12}$ & 0.534 \\
U.S.\ Open women  & SQS$_1$ & 85 & 0.133 & $2.7\times10^{-9}$ & 0.387 \\
U.S.\ Open women  & wElo    & 85 & 0.140 & $6.0\times10^{-17}$ & 0.528 \\
\midrule
\multicolumn{6}{l}{\textit{Second serve}} \\
\midrule
Wimbledon men   & SQS$_2$ & 47 & 0.020 & 0.586 & 0.257 \\
Wimbledon men   & wElo    & 47 & 0.071 & 0.018 & 0.170 \\
Wimbledon women & SQS$_2$ & 45 & $-0.017$ & 0.635 & 0.089 \\
Wimbledon women & wElo    & 45 & 0.107 & 0.005 & 0.525 \\
U.S.\ Open men    & SQS$_2$ & 78 & $-0.095$ & $5.9\times10^{-5}$ & $-0.358$ \\
U.S.\ Open men    & wElo    & 78 & 0.095 & $1.2\times10^{-7}$ & 0.384 \\
U.S.\ Open women  & SQS$_2$ & 66 & $-0.004$ & 0.873 & 0.000 \\
U.S.\ Open women  & wElo    & 66 & 0.091 & $1.3\times10^{-4}$ & 0.365 \\
\bottomrule
\end{widetabular}
\end{table}

The temporal results are broadly consistent with the within-year random split.
On first serves, SQS maintains stronger correlations with serve efficiency than wElo across all four datasets (Wimbledon men: $r=0.649$ vs.\ $0.010$; Wimbledon women: $r=0.365$ vs.\ $0.087$; U.S.\ Open men: $r=0.297$ vs.\ $0.247$; U.S.\ Open women: $r=0.578$ vs.\ $0.459$).
On second serves, SQS--efficiency associations are positive in three of four datasets, mirroring the pattern observed under random splitting.
Win percentage associations are weaker and more variable under the temporal split as well, consistent with the main analysis: SQS is most informative for the serve-proximal outcome, while overall point wins depend increasingly on non-serve factors.
These findings indicate that the serve profiles captured by SQS are stable across seasons and not merely artifacts of within-year overfitting.
}\normalcolor

\clearpage

{\color{black}%
\section{Point-Level Feature Robustness Check}\label{app:point_level}
The main model in Section~\ref{sec:methods} constructs SQS using \textit{server-level} summaries of serve behavior, such as average speed, speed variability, modal location, and location entropy. While this choice yields interpretable player-level profiles, it compresses within-player variation across points. To assess whether the main results depend on this aggregation, we estimate an alternative set of GLMMs using \textit{point-level} serve features directly.

For each serve type, the alternative model includes point-level serve speed, point-level serve location bin, and an indicator for whether the observed serve location matches the player's modal location for that serve type. As in the main model, we include crossed random intercepts for the server and returner. For the point-level mixed-effects model, let \(Y_{i,j}\) denote the binary indicator that server \(j\) wins point \(i\) within three shots. We then model:
\[
\text{logit}\Pr\left(Y_{i,j}=1\right) = \beta_0^{(p)} + \beta_1^{(p)} \text{speed}^{(z)}_{i,j} + (\boldsymbol{\beta}^{(p)}_{2})^{\top} \mathbb{I}(\text{loc}_{i,j}) + \beta_3^{(p)} \mathbb{I}(\text{modal\_match}_{i,j}) + u_j^{(p)} + v_k^{(p)}.
\]
Here, \(\text{speed}^{(z)}_{i,j}\) is the standardized serve speed on point \(i\), \(\mathbb{I}(\text{loc}_{i,j})\) is a one-hot encoding of the serve's location bin, and \(\mathbb{I}(\text{modal\_match}_{i,j})\) indicates whether the observed location matches server \(j\)'s modal location for that serve type. The terms \(u_j^{(p)}\) and \(v_k^{(p)}\) are server- and returner-specific random intercepts, respectively. We split the data between first and second serves for these mixed-effects models.

From the fitted models, we then construct a point-level version of SQS, denoted $\text{SQS}_s^{(p)}$ for serve type $s \in \{1, 2\}$. These scores are calculated by combining each player’s estimated fixed and random effects for each serve type. We evaluate $\text{SQS}_s^{(p)}$ out of sample using the same testing framework as in the main analysis (with an 80/20 random split within each year). 
\[
\text{SQS}^{(p)}_{j} = \hat{\beta}^{(p)}_{0} + \hat{\beta}^{(p)}_{1} \overline{\text{speed}^{(z)}_{j}} + (\hat{\boldsymbol{\beta}}^{(p)}_{2})^{\top} \overline{\mathbb{I}(loc_{j})} + \hat{\beta}^{(p)}_{3} \overline{\mathbb{I}(\text{modal\_match}_{j})} + \hat{u}^{(p)}_{j}.
\]

The bars above the covariates indicate that we use the average point-level data across all serves of type $s$ for server $j$ in the training data. Thus, the point-level model is first estimated using individual serve observations, but the resulting SQS is evaluated at the average serve characteristics of each player. This differs from the main specification, where the regressors themselves are constructed directly as server-level summaries (e.g., average speed and location entropy).

Tables~\ref{tab:point_level_eff} and~\ref{tab:point_level_win} report out-of-sample results for serve efficiency and win percentage, respectively, alongside weighted Elo (wElo) as a benchmark. These results allow us to compare the predictive performance of the point-level specification with that of the main server-level model. 

\begin{table}[htbp!]
\centering\color{black}
\caption{Point-level feature robustness check: serve efficiency.}
\label{tab:point_level_eff}
\footnotesize
\begin{widetabular}{llcccc}
\toprule
\textbf{Dataset} & \textbf{Predictor} & \textbf{n} & \textbf{RMSE} & \textbf{$p$-value} & \textbf{Correlation ($r$)} \\
\midrule
\multicolumn{6}{l}{\textit{First serve}} \\
\midrule
Wimbledon men   & SQS$_1^{(p)}$ & 62 & 0.805 & $2.39\times 10^{-9}$ & 0.671 \\
Wimbledon men   & wElo    & 62 & 1.295 & 0.251 & 0.148 \\
Wimbledon women & SQS$_1^{(p)}$ & 65 & 0.865 & $3.59\times 10^{-8}$ & 0.620 \\
Wimbledon women & wElo    & 65 & 1.473 & 0.418 & $-0.102$ \\
U.S.\ Open men  & SQS$_1^{(p)}$ & 86 & 1.199 & 0.011 & 0.273 \\
U.S.\ Open men  & wElo    & 86 & 1.569 & 0.023 & $-0.245$ \\
U.S.\ Open women & SQS$_1^{(p)}$ & 92 & 1.249 & 0.043 & 0.212 \\
U.S.\ Open women & wElo    & 92 & 1.455 & 0.506 & $-0.070$ \\
\midrule
\multicolumn{6}{l}{\textit{Second serve}} \\
\midrule
Wimbledon men   & SQS$_2^{(p)}$ & 60 & 1.216 & 0.056 & 0.248 \\
Wimbledon men   & wElo    & 60 & 1.338 & 0.495 & 0.090 \\
Wimbledon women & SQS$_2^{(p)}$ & 45 & 1.118 & 0.015 & 0.360 \\
Wimbledon women & wElo    & 45 & 1.311 & 0.429 & 0.121 \\
U.S.\ Open men  & SQS$_2^{(p)}$ & 82 & 1.425 & 0.803 & $-0.028$ \\
U.S.\ Open men  & wElo    & 82 & 1.615 & 0.003 & $-0.321$ \\
U.S.\ Open women & SQS$_2^{(p)}$ & 63 & 1.274 & 0.170 & 0.175 \\
U.S.\ Open women & wElo    & 63 & 1.587 & 0.027 & $-0.279$ \\
\bottomrule
\end{widetabular}
\end{table}

\begin{table}[htbp!]
\centering\color{black}
\caption{Point-level feature robustness check: win percentage.}
\label{tab:point_level_win}
\footnotesize
\begin{widetabular}{llcccc}
\toprule
\textbf{Dataset} & \textbf{Predictor} & \textbf{n} & \textbf{RMSE} & \textbf{$p$-value} & \textbf{Correlation ($r$)} \\
\midrule
\multicolumn{6}{l}{\textit{First serve}} \\
\midrule
Wimbledon men   & SQS$_1^{(p)}$ & 62 & 1.151 & 0.010 & 0.327 \\
Wimbledon men   & wElo    & 62 & 1.304 & 0.293 & 0.136 \\
Wimbledon women & SQS$_1^{(p)}$ & 65 & 1.010 & $4.70\times 10^{-5}$ & 0.482 \\
Wimbledon women & wElo    & 65 & 1.487 & 0.331 & $-0.123$ \\
U.S.\ Open men  & SQS$_1^{(p)}$ & 86 & 1.302 & 0.191 & 0.142 \\
U.S.\ Open men  & wElo    & 86 & 1.398 & 0.918 & 0.011 \\
U.S.\ Open women & SQS$_1^{(p)}$ & 92 & 1.315 & 0.232 & 0.126 \\
U.S.\ Open women & wElo    & 92 & 1.384 & 0.763 & 0.032 \\
\midrule
\multicolumn{6}{l}{\textit{Second serve}} \\
\midrule
Wimbledon men   & SQS$_2^{(p)}$ & 60 & 1.573 & 0.047 & $-0.258$ \\
Wimbledon men   & wElo    & 60 & 1.363 & 0.673 & 0.056 \\
Wimbledon women & SQS$_2^{(p)}$ & 45 & 1.296 & 0.353 & 0.142 \\
Wimbledon women & wElo    & 45 & 1.180 & 0.056 & 0.287 \\
U.S.\ Open men  & SQS$_2^{(p)}$ & 82 & 1.456 & 0.516 & $-0.073$ \\
U.S.\ Open men  & wElo    & 82 & 1.282 & 0.132 & 0.168 \\
U.S.\ Open women & SQS$_2^{(p)}$ & 63 & 1.453 & 0.574 & $-0.072$ \\
U.S.\ Open women & wElo    & 63 & 1.450 & 0.599 & $-0.067$ \\
\bottomrule
\end{widetabular}
\end{table}

\newpage
Relative to the main model, the point-level feature model produces a similar overall pattern: $\text{SQS}^{(p)}$ remains more informative for the serve-proximal outcome than for overall win percentage. For first serves, point-level $\text{SQS}_1^{(p)}$ continues to outperform wElo in predicting serve efficiency across all four datasets with notably stronger correlations.
On second serves, the evidence is more mixed but still generally favors $\text{SQS}^{(p)}$ for serve efficiency in three of the four datasets, while U.S.\ Open men remains weak for both metrics.

For win percentage, the results are again weaker and less consistent, which matches the main analysis in Section~\ref{sec:results} and Appendix~\ref{app:temporal}.
In particular, the point-level specification does not significantly strengthen the association between $\text{SQS}^{(p)}$ and overall point outcomes. Taken together, these results suggest that incorporating point-level serve speed and location does not overturn the main conclusions of the paper. The aggregated server-level model from Section~\ref{sec:methods:glmm} seems to retain most of the predictive signal relevant to serve efficiency, while remaining easier to interpret as a stable player-level serve profile.
}\normalcolor

\clearpage

\section{Top Server Rankings by Tournament}\label{app:rankings}
These tables report top-10 server rankings by SQS within each tournament and serve type.
Values are centered by subtracting the tournament--serve-type mean SQS (log-odds).
Positive values indicate above-average serving performance within the tournament and serve type, while negative values indicate below-average performance. {\color{black} We also report 95\% confidence intervals for player SQS estimates---these intervals are calculated using the covariance matrix of the fixed effects and the conditional variance of the player-specific random intercepts from the mixed-effects models. These intervals also provide a practical ranking-stability diagnostic: when adjacent players' intervals overlap materially, interpretation should emphasize performance tiers rather than exact ordinal rank positions.

\begin{table}[htbp!]
\centering\color{black}
\caption{Wimbledon men's singles: top 10 SQS rankings (centered).}
\label{tab:rank_wimb_men}
\footnotesize
\setlength{\tabcolsep}{3.5pt}

\begin{minipage}[t]{0.48\textwidth}
\centering
\vspace{0pt}
\textbf{First serve}\\
\begin{tabular}{r l l l}
\toprule
\textbf{Rank} & \textbf{Server} & \textbf{SQS} & \textbf{95\% CI} \\
\midrule
1  & John Isner           & 0.522 & (0.388, 0.656) \\
2  & Milos Raonic         & 0.485 & (0.314, 0.656) \\
3  & Nicolas Jarry        & 0.473 & (0.263, 0.683) \\
4  & Tim Van Rijthoven    & 0.435 & (0.249, 0.620) \\
5  & Sam Querrey          & 0.409 & (0.222, 0.596) \\
6  & Quentin Halys        & 0.398 & (0.211, 0.586) \\
7  & Marin Cilic          & 0.380 & (0.207, 0.552) \\
8  & Matteo Berrettini    & 0.372 & (0.236, 0.509) \\
9  & Nick Kyrgios         & 0.372 & (0.217, 0.527) \\
10 & Kevin Anderson       & 0.370 & (0.222, 0.518) \\
\bottomrule
\end{tabular}
\end{minipage}
\hfill
\begin{minipage}[t]{0.48\textwidth}
\centering
\vspace{0pt}
\textbf{Second serve}\\
\begin{tabular}{r l l l}
\toprule
\textbf{Rank} & \textbf{Server} & \textbf{SQS} & \textbf{95\% CI} \\
\midrule
1  & Maxime Cressy       & 0.474 & (0.236, 0.712) \\
2  & Milos Raonic        & 0.325 & (0.121, 0.531) \\
3  & John Isner          & 0.310 & (0.113, 0.509) \\
4  & Tim Van Rijthoven   & 0.267 & (0.051, 0.484) \\
5  & Quentin Halys       & 0.255 & (0.037, 0.474) \\
6  & Jeremy Chardy       & 0.237 & (0.013, 0.462) \\
7  & Nick Kyrgios        & 0.235 & (0.023, 0.449) \\
8  & Kevin Anderson      & 0.232 & (0.036, 0.429) \\
9  & Brandon Nakashima   & 0.226 & (-0.232, 0.686) \\
10 & Bradley Klahn       & 0.218 & (-0.016, 0.454) \\
\bottomrule
\end{tabular}
\end{minipage}
\end{table}

\begin{table}[htbp!]
\centering\color{black}
\caption{Wimbledon women's singles: top 10 SQS rankings (centered).}
\label{tab:rank_wimb_women}
\footnotesize
\setlength{\tabcolsep}{3.5pt}

\begin{minipage}[t]{0.48\textwidth}
\centering
\vspace{0pt}
\textbf{First serve}\\
\resizebox{\linewidth}{!}{%
\begin{tabular}{r l l l}
\toprule
\textbf{Rank} & \textbf{Server} & \textbf{SQS} & \textbf{95\% CI} \\
\midrule
1  & Elena Rybakina         & 0.414 & (0.247, 0.581) \\
2  & Serena Williams        & 0.358 & (0.174, 0.544) \\
3  & Jaqueline Cristian     & 0.342 & (0.056, 0.630) \\
4  & Ekaterina Alexandrova  & 0.323 & (0.096, 0.551) \\
5  & Qinwen Zheng           & 0.322 & (0.114, 0.531) \\
6  & Julia Goerges          & 0.314 & (0.124, 0.505) \\
7  & Ashleigh Barty         & 0.289 & (0.127, 0.452) \\
8  & Aryna Sabalenka        & 0.268 & (0.086, 0.450) \\
9  & Donna Vekic            & 0.266 & (0.089, 0.443) \\
10 & Stefanie Voegele       & 0.265 & (0.037, 0.494) \\
\bottomrule
\end{tabular}
}
\end{minipage}
\hfill
\begin{minipage}[t]{0.48\textwidth}
\centering
\vspace{0pt}
\textbf{Second serve}\\
\resizebox{\linewidth}{!}{%
\begin{tabular}{r l l l}
\toprule
\textbf{Rank} & \textbf{Server} & \textbf{SQS} & \textbf{95\% CI} \\
\midrule
1  & Camila Giorgi         & 0.278 & (0.138, 0.418) \\
2  & Johanna Konta         & 0.204 & (-0.062, 0.470) \\
3  & Sofia Kenin           & 0.162 & (-0.103, 0.429) \\
4  & Dayana Yastremska     & 0.159 & (0.069, 0.250) \\
5  & Aryna Sabalenka       & 0.156 & (0.044, 0.268) \\
6  & Petra Kvitova         & 0.153 & (0.038, 0.269) \\
7  & Danielle Collins      & 0.151 & (0.054, 0.248) \\
8  & Julia Goerges         & 0.150 & (0.047, 0.254) \\
9  & Cori Gauff            & 0.148 & (-0.003, 0.300) \\
10 & Clara Burel           & 0.145 & (-0.045, 0.335) \\
\bottomrule
\end{tabular}
}
\end{minipage}
\end{table}

\begin{table}[htbp!]
\centering\color{black}
\caption{U.S.\ Open men's singles: top 10 SQS rankings (centered).}
\label{tab:rank_us_men}
\footnotesize
\setlength{\tabcolsep}{3.5pt}

\begin{minipage}[t]{0.48\textwidth}
\centering
\vspace{0pt}
\textbf{First serve}\\
\resizebox{\linewidth}{!}{%
\begin{tabular}{r l l l}
\toprule
\textbf{Rank} & \textbf{Server} & \textbf{SQS} & \textbf{95\% CI} \\
\midrule
1  & Bradley Klahn        & 0.498 & (0.263, 0.733) \\
2  & Sam Querrey          & 0.477 & (0.253, 0.701) \\
3  & Gianluca Mager       & 0.456 & (0.215, 0.697) \\
4  & John Isner           & 0.456 & (0.287, 0.625) \\
5  & Milos Raonic         & 0.424 & (0.204, 0.645) \\
6  & Kevin Anderson       & 0.419 & (0.230, 0.607) \\
7  & Nick Kyrgios         & 0.361 & (0.183, 0.539) \\
8  & Reilly Opelka        & 0.351 & (0.154, 0.546) \\
9  & Jiri Vesely          & 0.339 & (0.130, 0.548) \\
10 & Alexander Bublik     & 0.312 & (0.093, 0.531) \\
\bottomrule
\end{tabular}
}
\end{minipage}
\hfill
\begin{minipage}[t]{0.48\textwidth}
\centering
\vspace{0pt}
\textbf{Second serve}\\
\resizebox{\linewidth}{!}{%
\begin{tabular}{r l l l}
\toprule
\textbf{Rank} & \textbf{Server} & \textbf{SQS} & \textbf{95\% CI} \\
\midrule
1  & Carlos Taberner      & 0.451 & (0.077, 0.825) \\
2  & Jiri Lehecka         & 0.361 & (-0.062, 0.784) \\
3  & Stefano Travaglia    & 0.346 & (-0.046, 0.738) \\
4  & James Duckworth      & 0.325 & (0.016, 0.633) \\
5  & Vasek Pospisil       & 0.324 & (-0.018, 0.666) \\
6  & Jiri Vesely          & 0.315 & (-0.066, 0.695) \\
7  & Denis Kudla          & 0.312 & (-0.001, 0.625) \\
8  & John Isner           & 0.308 & (-0.026, 0.641) \\
9  & Alexei Popyrin       & 0.276 & (-0.029, 0.581) \\
10 & Cristian Garin       & 0.268 & (-0.051, 0.586) \\
\bottomrule
\end{tabular}
}
\end{minipage}
\end{table}

\begin{table}[htbp!]
\centering\color{black}
\caption{U.S.\ Open women's singles: top 10 SQS rankings (centered).}
\label{tab:rank_us_women}
\footnotesize
\setlength{\tabcolsep}{3.5pt}

\begin{minipage}[t]{0.48\textwidth}
\centering
\vspace{0pt}
\textbf{First serve}\\
\resizebox{\linewidth}{!}{%
\begin{tabular}{r l l l}
\toprule
\textbf{Rank} & \textbf{Server} & \textbf{SQS} & \textbf{95\% CI} \\
\midrule
1  & Samantha Stosur       & 0.620 & (0.348, 0.892) \\
2  & Serena Williams       & 0.517 & (0.323, 0.712) \\
3  & Elena Rybakina        & 0.414 & (0.212, 0.615) \\
4  & Jodie Burrage         & 0.410 & (0.178, 0.643) \\
5  & Qinwen Zheng          & 0.410 & (0.230, 0.589) \\
6  & Liudmila Samsonova    & 0.393 & (0.196, 0.590) \\
7  & Julia Goerges         & 0.387 & (0.184, 0.589) \\
8  & Donna Vekic           & 0.383 & (0.201, 0.564) \\
9  & Diane Parry           & 0.343 & (0.111, 0.575) \\
10 & Rebeka Masarova       & 0.342 & (0.101, 0.583) \\
\bottomrule
\end{tabular}
}
\end{minipage}
\hfill
\begin{minipage}[t]{0.48\textwidth}
\centering
\vspace{0pt}
\textbf{Second serve}\\
\resizebox{\linewidth}{!}{%
\begin{tabular}{r l l l}
\toprule
\textbf{Rank} & \textbf{Server} & \textbf{SQS} & \textbf{95\% CI} \\
\midrule
1  & Liudmila Samsonova    & 0.482 & (0.182, 0.782) \\
2  & Ashlyn Krueger        & 0.441 & (0.038, 0.845) \\
3  & Barbora Krejcikova    & 0.382 & (0.115, 0.649) \\
4  & Coco Vandeweghe       & 0.360 & (0.009, 0.712) \\
5  & Rebecca Peterson      & 0.315 & (-0.010, 0.641) \\
6  & Caroline Garcia       & 0.286 & (0.005, 0.568) \\
7  & Petra Kvitova         & 0.282 & (-0.031, 0.595) \\
8  & Danielle Collins      & 0.250 & (-0.023, 0.524) \\
9  & Rebeka Masarova       & 0.241 & (-0.133, 0.615) \\
10 & Serena Williams       & 0.236 & (-0.017, 0.490) \\
\bottomrule
\end{tabular}
}
\end{minipage}
\end{table}

}\normalcolor

\end{document}